\documentclass[onecolumn,showpacs]{revtex4}

\usepackage{graphicx}
\usepackage{dcolumn}
\usepackage{bm}

\input epsf

\begin{document}

\title{\Large  Variable Modified Chaplygin Gas and Accelerating Universe}

\author{\bf Ujjal Debnath\footnote{ujjaldebnath@yahoo.com , ujjal@iucaa.ernet.in}}

\affiliation{Department of Mathematics, Bengal Engineering and
Science University, Shibpur, Howrah-711 103, India.}

\date{\today}

\begin{abstract}
In this letter, I have proposed a model of variable modified
Chaplygin gas and shown its role in accelerating phase of the
universe. I have shown that the equation of state of this model
is valid from the radiation era to quiessence model. The graphical
representations of statefinder parameters characterize different
phase of evolution of the universe. All results presented in the
letter concerns the case $k=0$.
\end{abstract}

\pacs{}

\maketitle

Recent observations of the luminosity of type Ia supernovae
indicate [1, 2] an accelerated expansion of the universe and lead
to the search for a new type of matter which violate the strong
energy condition $\rho+3p<0$. The matter consent responsible for
such a condition to be satisfied at a certain stage of evolution
of the universe is referred to as {\it dark energy}. There are
different candidates to play the role of the dark energy. The type
of dark energy represented by a scalar field is often called {\it
Quintessence}. In particular one can try another type of dark
energy - the so-called {\it pure Chaplygin gas} which obeys an
equation of state like [3]
\begin{equation}
p=-B/\rho,
\end{equation}

 where $p$~ and $\rho$ are respectively the pressure and energy
density and $B$ is a positive constant. Subsequently the above
equation was modified to the form (known as {\it generalized
Chaplygin gas})

\begin{equation}
p=-B/\rho^{\alpha}~~ \text{with}~~ 0\le \alpha \le 1.
\end{equation}

This generalized model has been studied previously [4, 5]. There
are some works on {\it modified Chaplygin gas} obeying an equation
of state [6, 7]

\begin{equation}
p=A\rho-\frac{B}{\rho^{\alpha}} ~~~~\text{with}~~~~ 0\le \alpha
\le 1,~A,~B~\text{are ~positive~ constants}.
\end{equation}

This equation of state shows radiation era (when $A=1/3$) at one
extreme (when the scale factor $a(t)$ is vanishingly small) while
a $\Lambda$CDM model at the other extreme (when the scale factor
$a(t)$ is infinitely large). Guo and Jhang [8] first proposed {\it
variable Chaplygin gas} model with equation of state (1), where
$B$ is a positive function of the cosmological scale factor `$a$'
i.e., $B=B(a)$. This assumption is reasonable since $B(a)$ is
related to the scalar potential if we take the Chaplygin gas as a
Born-Infeld scalar field [9]. Later there are some works on
variable Chaplygin gas model [10].\\

The metric of a homogeneous and isotropic universe in FRW model is

\begin{equation}
ds^{2}=dt^{2}-a^{2}(t)\left[\frac{dr^{2}}{1-kr^{2}}+r^{2}(d\theta^{2}+sin^{2}\theta
d\phi^{2})\right]
\end{equation}

where $a(t)$ is the scale factor and $k~(=0,\pm 1)$ is the
constant curvature of their spatial sections.\\

The Einstein field equations are

\begin{equation}
\frac{\dot{a}^{2}}{a^{2}}+\frac{k}{a^{2}}=\frac{1}{3}\rho
\end{equation}
and
\begin{equation}
\frac{\ddot{a}}{a}=-\frac{1}{6}(\rho+3p)
\end{equation}

where $\rho$ and $p$ are energy density and isotropic pressure
respectively (choosing $8\pi G=c=1$).\\

The energy conservation equation is

\begin{equation}
\dot{\rho}+3\frac{\dot{a}}{a}(\rho+p)=0
\end{equation}

Now, I have introduced {\it variable modified Chaplygin gas} with
equation of state (3) where $B$ is a positive function of the
cosmological scale factor `$a$' (i.e., $B=B(a)$) as

\begin{equation}
p=A\rho-\frac{B(a)}{\rho^{\alpha}} ~~~~\text{with}~~~~ 0\le \alpha
\le 1,~A ~\text{is ~constant}>0.
\end{equation}

At all stages it shows a mixture. Also in between there is also
one stage when the pressure vanishes and the matter content is
equivalent to a pure dust.\\

Now, for simplicity, assume $B(a)$ is of the form
\begin{equation}
B(a)=B_{0}a^{-n}
\end{equation}

where $B_{0}>0$ and $n$ are constants. Using equations (7), (8)
and (9), I have the solution of $\rho$ as

\begin{equation}
\rho=\left[\frac{3(1+\alpha)B_{0}}{\{3(1+\alpha)(1+A)-n\}}~\frac{1}{a^{n}}+\frac{C}{a^{3(1+A)(1+\alpha)}}
\right]^{\frac{1}{1+\alpha}}
\end{equation}

where $C>0$ is an arbitrary integration constant and
$3(1+A)(1+\alpha)>n$, for positivity of first term. Here $n$ must
be positive, because otherwise, $a\rightarrow\infty$ implies $\rho\rightarrow\infty$,
which is not the case for expanding Universe.\\

From equation (5), for $k=0$, we get the explicit form of $t$ in
terms of $a$ as

$$
t=K
a^{\frac{n}{2(1+\alpha)}}~_{2}F_{1}[\frac{1}{2(1+\alpha)},-z,1-z,-\frac{C}{K}~a^{-\frac{n}{2(1+\alpha)z}}]
$$
where
$$
K=\frac{2}{n}~\left[(1+\alpha)^{\alpha}\sqrt{\frac{n}{6B_{0}z}}\right]^{\frac{1}{1+\alpha}}
~~,~z=\frac{n}{2(1+\alpha)\{3(1+A)(1+\alpha)-n\}}
$$

The deceleration parameter $q$ has the expression
$$q=-\frac{\ddot{a}}{aH^{2}}$$

For accelerating universe, $q$ must be negative i.e., $\ddot{a}>0$
i.e.,
\begin{equation}
\frac{2(1+\alpha)-n}{3(1+\alpha)(1+A)-n}~a^{3(1+\alpha)(1+A)-n}>\frac{C(1+3A)}{3B_{0}}
\end{equation}

which requires $n<2(1+\alpha)$. Since $0\le \alpha \le 1$, this implies $n\le 4$.\\

This expression shows that for small value of scale factor we have
decelerating universe while for large values of scale factor we
have accelerating universe and the transition occurs when the
scale factor has the expression
$a=\left[\frac{C(1+3A)\{3(1+\alpha)(1+A)-n
\}}{3B_{0}\{2(1+\alpha)-n\}}
\right]^{\frac{1}{3(1+\alpha)(1+A)-n}}$.\\

Now for small value of scale factor $a(t)$, I have

\begin{equation}
\rho\simeq \frac{C^{\frac{1}{1+\alpha}}}{a^{3(1+A)}}~,
\end{equation}

which is very large and corresponds to the universe dominated by
an equation of state $p=A\rho$.\\

Also for large value of scale factor $a(t)$,

\begin{equation}
\rho\simeq
\left(\frac{3(1+\alpha)B_{0}}{\{3(1+\alpha)(1+A)-n\}}\right)^{\frac{1}{1+\alpha}}~a^{-\frac{n}{1+\alpha}}
\end{equation}
and I obtain
\begin{equation}
p=\left(-1+\frac{n}{3(1+\alpha)}\right)\rho
\end{equation}

which correspond to {\it quiessence} model (i.e., dark energy with
constant equation of state) [13].\\

Note that $n = 0$ corresponds to the original modified Chaplygin
gas scenario [7], in which the modified Chaplygin gas behaves
initially radiation and later as a cosmological constant. However,
equation (10) shows that, in the variable modified Chaplygin gas
scenario, it interpolates between a radiation dominated phase
($A=1/3$) and a quiessence-dominated phase described by the
constant equation of state $p=\gamma \rho$ where $\gamma =
-1+\frac{n}{3(1+\alpha)}~<-\frac{1}{3}$ .\\

I have described this particular cosmological model from the field
theoretical point of view by introducing a scalar field $\phi$ and
a self-interacting potential $V(\phi)$ with the effective
Lagrangian

\begin{equation}
{\cal L_{\phi}}=\frac{1}{2}\dot{\phi}^{2}-V(\phi)
\end{equation}

In the paper of Gorini et al [4], it has been shown that for the
simple flat Friedmann model with Chaplygin gas can equivalently
described in terms of a homogeneous minimally coupled scalar field
$\phi$. Following Barrow [11], Kamenshchik et al [3, 12] have
obtained homogeneous scalar field $\phi(t)$ and a
potential $V(\phi)$ to describe Chaplygin cosmology.\\

Now, I consider the energy density and pressure corresponding to a
scalar field $\phi$ having a self-interacting potential $V(\phi)$.
The analogous energy density $\rho_{\phi}$ and pressure $p_{\phi}$
for the scalar field are the following:
\begin{equation}
\rho_{\phi}=\frac{1}{2}\dot{\phi}^{2}+V(\phi)=\rho=\left[\frac{3(1+\alpha)B_{0}}
{\{3(1+\alpha)(1+A)-n\}}~\frac{1}{a^{n}}+\frac{C}{a^{3(1+A)(1+\alpha)}}
\right]^{\frac{1}{1+\alpha}}
\end{equation}
and
\begin{eqnarray*}
p_{\phi}=\frac{1}{2}\dot{\phi}^{2}-V(\phi)=A\rho-\frac{B_{0}}{\rho^{\alpha}}
=A\left[\frac{3(1+\alpha)B_{0}}{\{3(1+\alpha)(1+A)-n\}}~\frac{1}{a^{n}}+\frac{C}{a^{3(1+A)(1+\alpha)}}
\right]^{\frac{1}{1+\alpha}}
\end{eqnarray*}

\begin{equation}
~~~~~~~~~~~~~~~~~~~~~~~~~~~~~~~~~~~~~
-B_{0}a^{-n}\left[\frac{3(1+\alpha)B_{0}}{\{3(1+\alpha)(1+A)-n\}}~\frac{1}{a^{n}}+\frac{C}{a^{3(1+A)(1+\alpha)}}
\right]^{-\frac{\alpha}{1+\alpha}}
\end{equation}
\\
Hence for flat universe (i.e., $k=0$), I have

\begin{equation}
\phi=\frac{\sqrt{1+A}}{\{3(1+\alpha)(1+A)-n\}}~\left[2~\text{log}(\sqrt{u+x}+\sqrt{u+y})
-\frac{\sqrt{x}}{\sqrt{y}}~\text{log}\left(
\frac{\left(\sqrt{x(u+y)}+\sqrt{y(u+x)}\right)^{2}}{x^{3/2}\sqrt{y}~u}
\right) \right]
\end{equation}

and
\begin{eqnarray*}
V(\phi)=\frac{1}{2}(1-A)\left[\frac{3(1+\alpha)B_{0}}{\{3(1+\alpha)(1+A)-n\}}~\frac{1}{a^{n}}+\frac{C}{a^{3(1+A)(1+\alpha)}}
\right]^{\frac{1}{1+\alpha}}
\end{eqnarray*}

\begin{equation}
~~~~~~~~~~~~~~~~~~~~~~~
+\frac{1}{2}B_{0}a^{-n}\left[\frac{3(1+\alpha)B_{0}}{\{3(1+\alpha)(1+A)-n\}}~\frac{1}{a^{n}}+\frac{C}{a^{3(1+A)(1+\alpha)}}
\right]^{-\frac{\alpha}{1+\alpha}}
\end{equation}

where $x=\frac{n}{1+A}$, $y=3(1+\alpha)$ and $u=\frac{nC}{B_{0}}
\left(\frac{y}{x}-1\right)~a^{n\left(1-\frac{y}{x}\right)}$.\\

\begin{figure}
\includegraphics[height=1.7in]{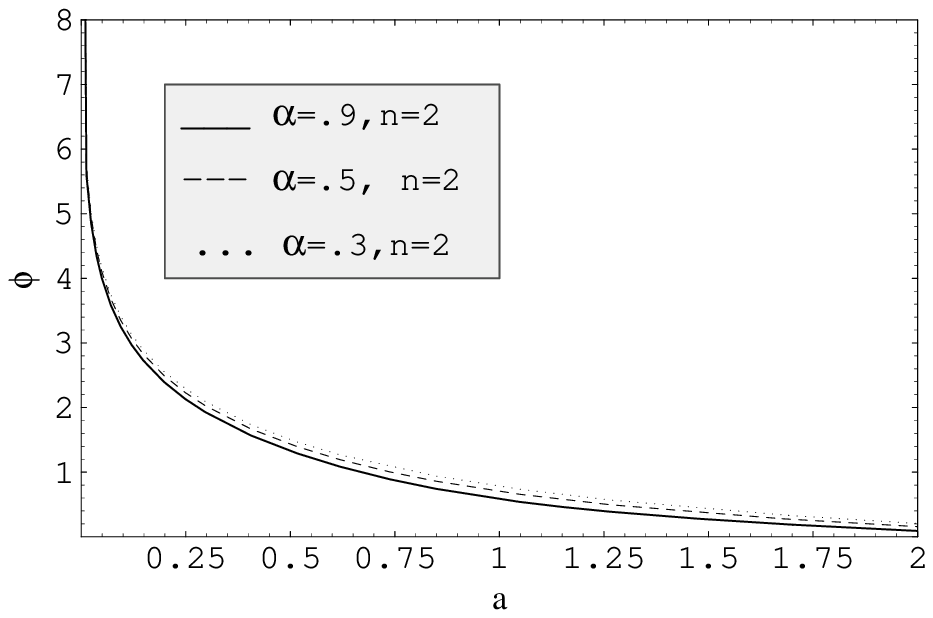}~~~
\includegraphics[height=1.7in]{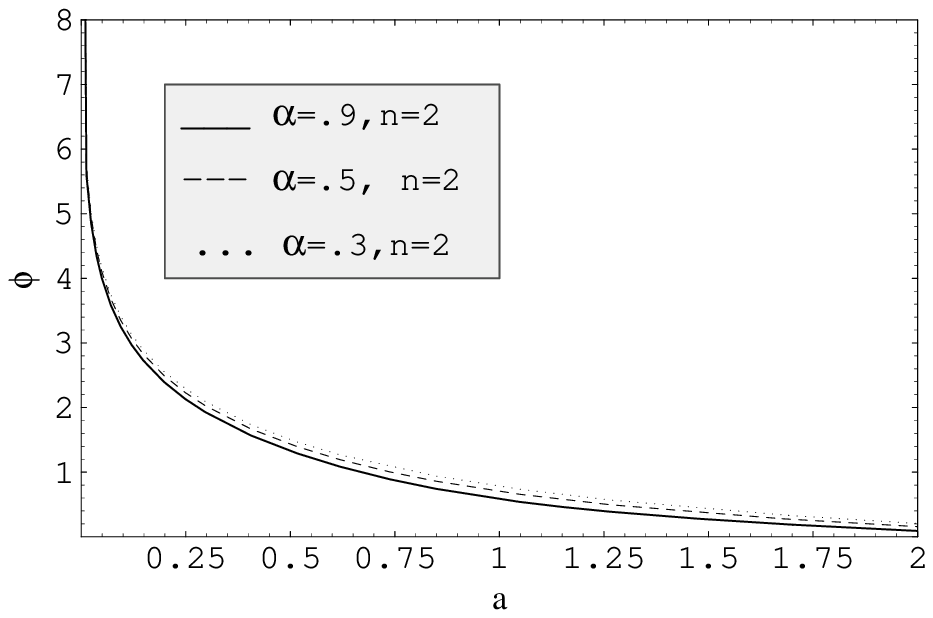}\\
\vspace{1mm}
Fig.1~~~~~~~~~~~~~~~~~~~~~~~~~~~~~~~~~~~~~~~~~~~~~~~~~~~~~Fig.2\\
\vspace{5mm} Fig. 1 shows variation of $\phi$ against $a$ for
$\alpha=0.6$ and various values of $n$ (= 1, 2, 3). Fig. 2 shows
variation of $\phi$ against $a$ for $n=2$ and various values of
$\alpha$ (= 0.9, 0.5., 0.3). \hspace{2cm} \vspace{6mm}

\includegraphics[height=1.7in]{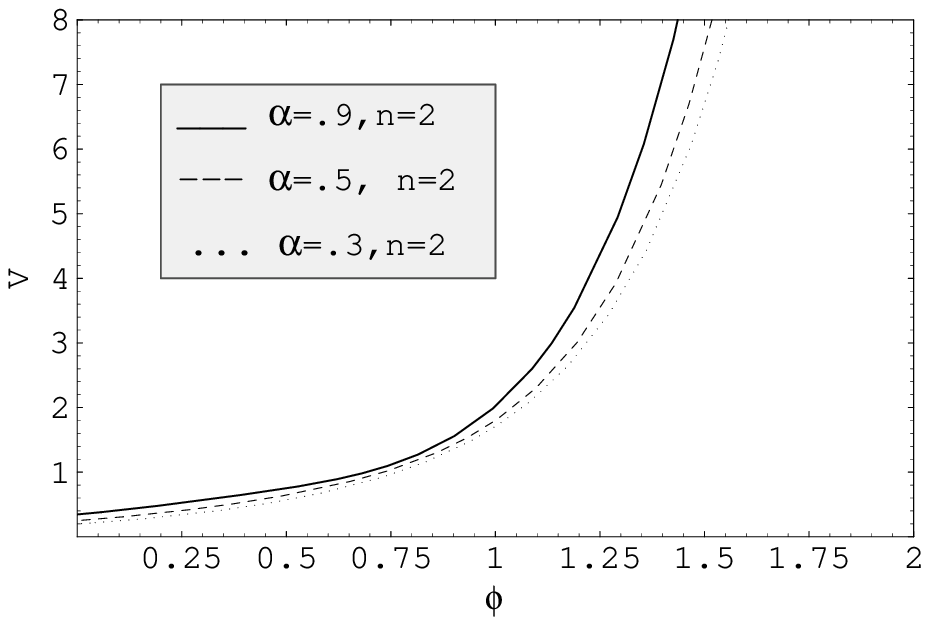}~~~
\includegraphics[height=1.7in]{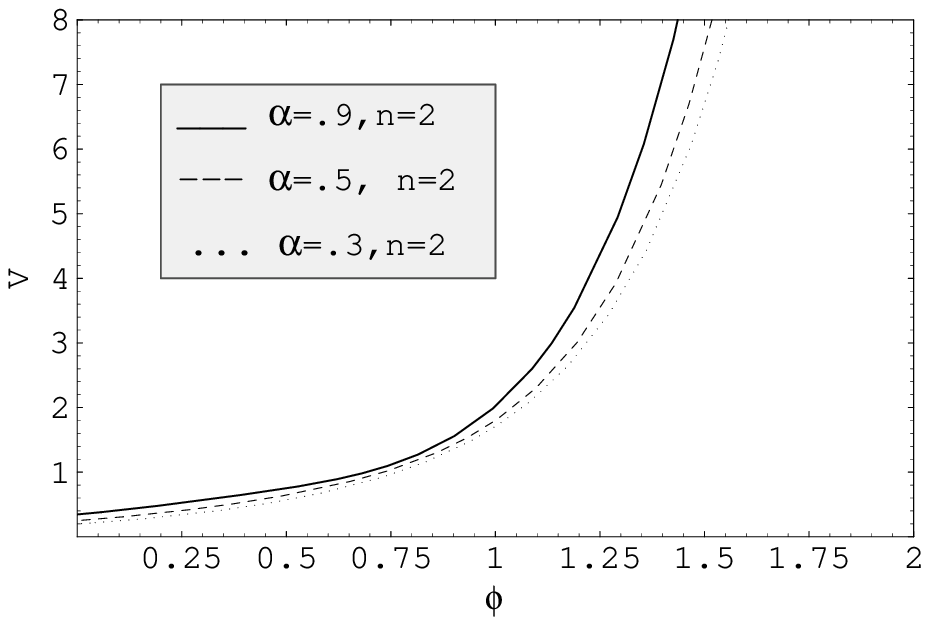}\\
\vspace{1mm}
Fig.3~~~~~~~~~~~~~~~~~~~~~~~~~~~~~~~~~~~~~~~~~~~~~~~~~~~~~Fig.4\\
\vspace{5mm} Fig. 3 shows variation of $V(\phi)$ against $\phi$
for $\alpha=0.6$ and various values of $n$ (= 1, 2, 3). Fig. 4
shows variation of $V(\phi)$ against $\phi$ for $n=2$ and various
values of $\alpha$ (= 0.9, 0.5., 0.3). \hspace{2cm} \vspace{6mm}

\end{figure}

The graphical representations of $\phi$ against $a$ and $V(\phi)$
against $\phi$ have been shown in figures 1, 2 and  figures 3, 4
respectively for $A=1/3$. Figures 1 and 3 show the fixed value of
$\alpha=0.6$ and various values of $n$ (= 1, 2, 3). In this case,
$V(\phi)$ increases with $\phi$ and slope of the curves decreases
as $n$ increases. Figures 2 and 4 show the fixed value of  $n=2$
and various values of $\alpha$ (= 0.9, 0.5., 0.3). In this case
also, $V(\phi)$ increases with $\phi$ and slope of the curves
decreases as $\alpha$ decreases. Figures 1 and 2 describe the
scalar field $\phi$ always decreases with the evolution of the universe.\\

In the paper [4], the flat Friedmann model filled with Chaplygin
fluid has been analyzed in terms of the recently proposed ``{\it
statefinder}'' parameters [14]. The statefinder diagnostic along
with future SNAP observations may perhaps be used to discriminate
between different dark energy models. The above statefinder
diagnostic pair i.e., $\{r,s\}$ parameters are constructed from
the scale factor $a(t)$ and its derivatives upto the third order
as follows:

\begin{equation}
r=\frac{\dddot{a}}{aH^{3}}~~~~\text{and}~~~~s=\frac{r-1}{3\left(q-\frac{1}{2}\right)}
\end{equation}

where $H$ is the Hubble parameter and
$q~\left(=-\frac{a\ddot{a}}{\dot{a}^{2}}\right)$ is the
deceleration parameter. These parameters are dimensionless and
allow us to characterize the properties of dark energy in a model
independent manner. The statefinder is dimensionless and is
constructed from the scale factor of the Universe and its time
derivatives only. The parameter $r$ forms the next step in the
hierarchy of geometrical
cosmological parameters after $H$ and $q$.\\

\begin{figure}
\includegraphics[height=2.7in]{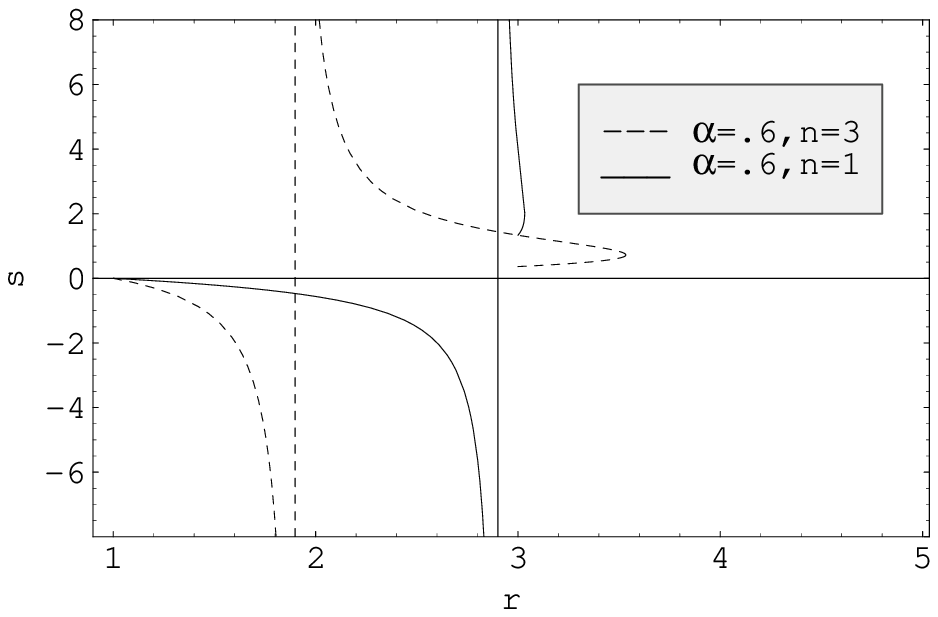}\\
\vspace{1mm} Fig.5\\

\vspace{5mm} Fig. 5 shows the variation of $s$
 against $r$ for different values of $n$ (= 3, 1) and for $\alpha=0.6$, $A=1/3$.\hspace{2cm} \vspace{4mm}

\end{figure}

For Friedmann model with flat universe (i.e., $k=0$),

\begin{equation}
H^{2}=\frac{\dot{a}^{2}}{a^{2}}=\frac{1}{3}\rho
\end{equation}
and
\begin{equation}
q=-\frac{\ddot{a}}{aH^{2}}=\frac{1}{2}+\frac{3}{2}\frac{p}{\rho}
\end{equation}

So from equation (20) we get

\begin{equation}
r=1+\frac{9}{2}\left(1+\frac{p}{\rho}\right)\frac{\partial
p}{\partial\rho}~,~~~~~s=\left(1+\frac{\rho}{p}\right)\frac{\partial
p}{\partial\rho}
\end{equation}

Thus, I get the ratio between $p$ and $\rho$:

\begin{equation}
\frac{p}{\rho}=\frac{2(r-1)}{9s}
\end{equation}

For variable modified Chaplygin gas, using equations (7), (23) and
(24), I get the relation between $r$ and $s$:

\begin{equation}
18(r-1)s^{2}+18\alpha s(r-1)+4\alpha
(r-1)^{2}=9sA(1+\alpha)(2r+9s-2)+3ns(2r-2-9sA)
\end{equation}

Figure 5 shows the variation of $s$ with the variation of $r$ for
$A=1/3$ and for $\alpha=~0.6$ and $n=~3,~1$. The portion of the
curve on the positive side of $s$ which is physically admissible
is only the values of $r$ greater than
$\left\{1+\frac{9}{2}A(1+A)\right\}$. The part of the curve
between $r=1$ and $r=1+\frac{9}{2}A(1+A)$ with positive value of
$s$ is not admissible (we have not shown that part in the figure
5) because for the Chaplygin gas under consideration we face a
situation, where the magnitude of the constant $B_{0}$ becomes
negative. \\

Thus the curve in the positive side of $s$ starts from radiation
era and goes asymptotically to the dust model. But the portion in
the negative side of $s$ represents the evolution from dust state
($s=-\infty$) to the quiessence model. Thus the total curve
represents the evolution of the universe starting from the
radiation era to the quiessence model.\\

In this work, I have presented a model for variable modified
Chaplygin gas. In this model, I am able to describe the universe
from the radiation era ($A=1/3$ and $\rho$ is very large) to
quiessence model ($\rho$ is small constant). So compare to
Chaplygin gas model, the present model describe universe to a
large extent. Also if, I  put $A=0$ with $\alpha=1$, then I can
recover the results of the Chaplygin gas model. If put $n=0$ then
variable modified Chaplygin gas model reduces to only modified
Chaplygin gas model [7]. In figure 5, for $\{r,s\}$ diagram the
portion of the curve for $s>0$ between $r=1$ to
$r=1+\frac{9}{2}A(1+A)$ is not describable by the modified
Chaplygin gas under consideration. For example, if I choose
$r=1.03,~A=1/3$ then from the curve $s=0.01$ which corresponds to
$q=3/2$ and hence I have from the equation of state, $B<0$ which
is not valid for the specific Chaplygin gas model considered here.
At the large value of the scale factor I must have some stage
where the pressure becomes negative and hence $B$ has to be chosen
positive. It follows therefore that a portion of the curve
as mentioned above should not remain valid. \\\\

{\bf Acknowledgement:}\\

The author would like to express his gratitude to the authority of
IUCAA, Pune for providing him the Associateship Programme under
which a part of this work was carried out. \\

{\bf References:}\\
\\
$[1]$  N. A. Bachall, J. P. Ostriker, S. Perlmutter and P. J.
Steinhardt, {\it Science} {\bf 284} 1481 (1999).\\
$[2]$ S. J. Perlmutter et al, {\it Astrophys. J.} {\bf 517} 565
(1999).\\
$[3]$ A. Kamenshchik, U. Moschella and V. Pasquier, {\it Phys.
Lett. B} {\bf 511} 265 (2001); V. Gorini, A. Kamenshchik, U.
Moschella and V. Pasquier, {\it gr-qc}/0403062.\\
$[4]$ V. Gorini, A. Kamenshchik and U. Moschella, {\it Phys. Rev.
D} {\bf 67} 063509 (2003); U. Alam, V. Sahni , T. D. Saini and
A.A. Starobinsky, {\it Mon. Not. Roy. Astron. Soc.} {\bf 344}, 1057 (2003).\\
$[5]$ M. C. Bento, O. Bertolami and A. A. Sen, {\it Phys. Rev. D}
{66} 043507 (2002).\\
$[6]$ H. B. Benaoum, {\it hep-th}/0205140.\\
$[7]$ U. Debnath, A. Banerjee and S. Chakraborty, {\it Class.
Quantum Grav.} {\bf 21} 5609 (2004).\\
$[8]$ Z. K. Guo and Y. Z. Zhang, {\it Phys. Lett. B} {\bf 645} 326 (2007); {\it astro-ph}/0506091.\\
$[9]$ M.C. Bento, O. Bertolami and A.A. Sen, {\it Phys. Lett. B}
{\bf 575} 172 (2003).\\
$[10]$ G. Sethi, S. K. Singh, P. Kumar, D. Jain and A. Dev, {\it
Int. J. Mod. Phys. D} {\bf 15} 1089 (2006); {\it
astro-ph}/0508491; Z. K. Guo and Y. Z. Zhang, {\it
astro-ph}/0509790.\\
$[11]$ J. D. Barrow, {\it Nucl. Phys. B} {\bf 310} 743 (1988); {\it Phys.Lett. B} {\bf 235} 40 (1990).\\
$[12]$ V. Gorini, A. Yu. Kamenshchik, U. Moschella, V. Pasquier,
{\it Phys.Rev. D} {\bf 69} 123512 (2004).\\
$[13]$ Z.K. Guo, N. Ohta and Y.Z. Zhang, {\it astro-ph}/0505253.\\
$[14]$ V. Sahni, T. D. Saini, A. A. Starobinsky and U. Alam, {\it
JETP Lett.} {\bf 77} 201 (2003).\\

\end{document}